\documentclass[prd,twocolumn,nofootinbib,superscriptaddress]{revtex4-2}
\input epsf
\usepackage{graphics}
\usepackage{amsmath,amssymb}
\usepackage{color}
\usepackage{dcolumn}
\usepackage{hyphenat}
\usepackage{multirow}

\usepackage{bm}

\def\be{\begin{equation}}
\def\ee{\end{equation}}
\def\ba{\begin{eqnarray}}
\def\ea{\end{eqnarray}}

\usepackage{graphicx}

\begin{document}

\title{Cosmic strings in the complex symmetron model}

\author{Ali Nezhadsafavi} 
\email[]{ali\_nezhadsafavi@sfu.ca}
\affiliation{Department of Physics, Simon Fraser University, Burnaby, BC, V5A 1S6, Canada}
\author{Levon Pogosian} 
\email[]{levon\_pogosian@sfu.ca}
\affiliation{Department of Physics, Simon Fraser University, Burnaby, BC, V5A 1S6, Canada}

\begin{abstract}
We study cosmic strings in the complex symmetron model, a scalar-tensor theory with a spontaneously broken local $U(1)$ symmetry in low matter density regions. Using numerical simulations, we show that these strings preferentially attach to matter halos, leading to the stabilization of string loops. While the requirement for screening of fifth-force interactions in the solar system limits observable signatures in theories with universal coupling to matter, analogous topological defects in the dark sector may still influence cosmic structure formation, offering a novel avenue to constrain dark-sector interactions.
\end{abstract}

\maketitle
\section{Introduction}

The presence of topological defect configurations in a physical system, whether in condensed matter or particle field theory, can significantly influence its macroscopic behavior and give rise to distinctive phenomenological signatures. Most studies of field-theoretic topological defects, such as magnetic monopoles, cosmic strings, and domain walls~\cite{Vilenkin:2000jqa}, have been conducted in models with minimal coupling to gravity, where the fields forming the defects interact gravitationally with other matter only through the impact of their stress-energy tensor on the metric $g_{\mu\nu}$. In contrast, non-minimally coupled field theories feature an additional (``fifth force'') interaction mediated by the same field that supports the defect, which also occurs in theories with explicit couplings between the scalar field and the Ricci scalar.

The discovery of cosmic acceleration~\cite{Perlmutter:1999qho,Riess:1998af} and the unresolved nature of dark matter have spurred interest in such non-minimally coupled theories, including coupled dark energy models, scalar-tensor theories, and other modified gravity scenarios. Given that terrestrial and solar-system gravity tests place tight constraints on deviations from General Relativity (GR), any fifth-force interactions must either be restricted to the dark sector or screened in high-density environments. Screening mechanisms such as the chameleon~\cite{Khoury:2003aq}, symmetron~\cite{Hinterbichler:2010es}, Vainshtein~\cite{Vainshtein:1972sx}, and k-mouflage~\cite{Babichev:2009ee} models enable modified gravity theories to evade local tests while exhibiting cosmologically relevant behavior. Over the past two decades, these mechanisms have been categorized broadly~\cite{Joyce:2014kja} based on whether the fifth-force coupling or range is suppressed in high-density regions, or if the force law changes dynamically due to derivative terms in the action.

The symmetron model is a scalar-tensor theory in which the coupling between the scalar field and matter vanishes in dense environments, thereby screening the fifth force in regions like the solar system. Although the model does not address the fine-tuning problems associated with the cosmological constant~\cite{Weinberg:1988cp,Padmanabhan:2002ji}, it remains a viable modified gravity framework with testable phenomenology~\cite{Brax:2012nk,Hojjati:2015ojt,Fischer:2024eic} and implications for cosmic structure formation. For example, in low-density regions, the spontaneously broken $\mathbb{Z}_2$ symmetry gives rise to domain walls, which have been observed in N-body simulations~\cite{Llinares:2013qbh} and further explored in~\cite{Llinares:2013jua}. Due to their coupling to matter, symmetron domain walls exhibit richer behavior than their minimally coupled counterparts, including attraction to regions of high density~\cite{Llinares:2014zxa,Pearson:2014vqa}, which allows closed walls to remain stable by pinning to the cosmic web. These domain walls have also been constrained in laboratory experiments~\cite{Clements:2023bva}, and symmetron fields have been proposed as dark matter candidates~\cite{Burrage:2018zuj,Kading:2023hdb}.

Motivated by the rich phenomenology of symmetron domain walls, we investigate cosmic string solutions in the complex field version of the symmetron model. Analogous to the domain wall case, we anticipate that cosmic strings in this model will preferentially attach to matter halos, lowering their energy and potentially stabilizing otherwise unstable configurations. Using numerical field theory simulations, we confirm this behavior and show that string loops can indeed be stabilized by attaching to multiple halos.

In what follows, we review the symmetron model and introduce its complex extension in Section~\ref{s:model}, describe the straight string and loop solutions and our simulation methods in Section~\ref{s:methods}, present the results in Section~\ref{s:results}, and conclude with a summary in Section~\ref{s:summary}.

\section{The Complex Symmetron Model}
\label{s:model}

We promote the real scalar field in the symmetron model \cite{Hinterbichler:2010es} to a complex field with a local $U(1)$ symmetry, and consider the action
\ba
\nonumber
S&=&\int d^{4}x \sqrt{-g }\left[R-\frac{1}{2}|D_{\mu}\phi|^{2}-V(|\phi|)-\frac{1}{4}F^{\mu\nu}F_{\mu\nu}\right] \\
&+&S_{m}(\tilde{g}_{\mu\nu},\psi),
\label{eq:action}
\ea
where the Ricci scalar $R$ is built from the Einstein frame metric $g_{\mu\nu}$, $\phi$ is a complex scalar field charged under the $U(1)$ gauge field $A_\mu$, $D_{\mu}\phi=\partial_{\mu}\phi-iqA_{\mu}\phi$ is the covariant derivative, $F^{\mu\nu}$ is the gauge field strength tensor, and $S_{m}(\tilde{g}_{\mu\nu},\psi)$ is the action for the matter fields (including dark matter) $\psi$ that follow the geodesics of the {\it Jordan frame} metric
\begin{equation}\label{eq:EJ}
\tilde{g}_{\mu\nu}=A^{2}(|\phi|)g_{\mu\nu}.
\end{equation}
This model features a non-minimal coupling the scalar field $\phi$ and matter, and is equivalent to a scalar-tensor modified gravity theory when reformulated in the Jordan frame. In this theory, non-relativistic matter experiences a fifth force mediated by $\phi$ and moves along geodesics of $\tilde{g}_{\mu\nu}$, whereas gravitational waves and relativistic fields propagate along geodesic of $g_{\mu\nu}$.

The coupling function $A^{2}(|\phi|)$ and potential $V(|\phi|)$ are chosen to ensure that the fifth force vanishes in regions of high matter density. This screening mechanism allows the theory to exhibit novel cosmological phenomenology while remaining consistent with the stringent solar system tests of GR. In the symmetron model, one has
\begin{equation}\label{eq:coupling}
A(|\phi|)=1+\frac{|\phi|^{2}}{2M},
\end{equation}
and
\ba
\nonumber
V(|\phi|)&=&-\frac{\mu^{2}}{2}|\phi|^{2}+\frac{\lambda}{4} |\phi|^{4}+V_{0} \\
&=& \frac{\lambda}{4}(|\phi|^{2}-\eta^{2})^{2} +V_{0} - \frac{\mu^4}{4\lambda},
\label{eq:potential}
\ea
where $\eta=\mu/\sqrt{\lambda}$ and $V_0$ is tuned so that the vacuum energy drives the observed cosmic acceleration. The equations of motion (EOM) are
\begin{align}
D^{\gamma}D_{\gamma} \phi-\lambda(\phi\phi^{\ast}-\eta^{2})\phi-\frac{\rho\phi}{M^{2}}&=0 \label{eq:EoM_1}\\
\frac{-iq}{2}\big[\phi^{\ast} D_{\beta}\phi-\phi (D_{\beta} \phi)^{\ast}\big]+\nabla^{\gamma} F_{\gamma\beta}&=0,
\label{eq:EoMs_2}
\end{align}
where $\rho$ is the energy density of the dust-like matter\footnote{Strictly speaking, $\phi$ couples to the trace of the matter stress-energy tensor, which vanishes for relativistic matter and equals $\rho$ for non-relativistic (dust-like) matter. In this paper, we treat all non-relativistic matter as dust.}. 

We can re-write Eq.~\eqref{eq:EoM_1} as 
\begin{equation}\label{eq:rewrite}
D^{\mu}D_{\mu}\phi=V_{\text{eff},\phi^{\ast}}
\end{equation}
where $V_{\text{eff}}$ is the effective potential given by
\ba
\nonumber
V_{\text{eff}}&=&\frac{\rho |\phi|^{2}}{2M^{2}}+\frac{\lambda}{4}(|\phi|^{2}-\eta^{2})^{2} +V_{0} - \frac{\mu^4}{4\lambda}  \\
&=& {1\over 2}(\frac{\rho}{M^{2}}-\mu^{2})|\phi|^{2}+\frac{\lambda}{4} |\phi|^4 +V_{0}.
\label{eq:effective_poten_vev}
\ea
This effective potential exhibits a spontaneously broken $U(1)$ symmetry when the matter density drops below a critical value $\rho_{c}=\mu^{2}M^{2}$. As a theory with a spontaneously broken axial symmetry, the complex symmetron model admits topologically stable line defects, {\it i.e.} cosmic strings\footnote{Cosmic string solutions also exist in the global $U(1)$ symmetry version of the symmetron. We focus on the local $U(1)$ case because the strings are more localized, making numerical simulations more efficient. The qualitative results would apply to global strings as well.}.

The model has four free parameters: $\mu$, which sets the mass of the scalar field; $\lambda$, which determines the depth of the potential; the gauge field coupling $q$; and $M$, which governs the coupling strength between matter and the scalar field.

The most notable feature of the symmetron model is the screening of fifth-force interactions in regions where $\rho > \rho_{c}$. To achieve screening near a mass halo, the mass distribution must satisfy the thin-shell condition~\cite{Joyce:2014kja}. For a spherical halo of density $\rho$ and radius $R$, the condition is satisfied when
\begin{equation}
\frac{\rho R^{2}}{M^{2}} \gg 1.
\label{eq:thin-shell}
\end{equation}
In what follows, we work with matter configurations that satisfy this condition.

\subsection{The Nielsen-Olesen symmetron string}

Cosmic strings are solutions of the field equations that are stable as a result of a non-trivial first homotopy group of the vacuum manifold~\cite{Vilenkin:2000jqa}, which is the case for the complex symmetron model with a spontaneously broken local $U(1)$ symmetry. The solution in cylindrical coordinates for a static straight cosmic string along along the $z$-axis is found by substituting the Nielsen-Olesen ansatz~\cite{Nielsen:1973cs},
\ba
\Phi&=&\eta f(r)e^{in\phi} \label{eq:str_scalar_ansatz} \\
A_\phi&=& \frac{n}{q}\alpha(r) \ ({\rm s.t.} \ {\vec A} = \hat{\phi}A_\phi/r), \ A_z=A_r=0,
\label{eq:str_gauge_ansatz}
\ea
into the EOM and solving for functions $f(r)$ and $\alpha(r)$ subject to boundary conditions $f(0)=\alpha(0)=0$, $f(r)\rightarrow1$, $\alpha(r)\rightarrow1$ as $r\rightarrow \infty$, where $n$ is the string winding number. The widths of the scalar and gauge cores of the string, $r_s$ and $r_v$, depend on the masses of the scalar and the gauge field~\cite{Vilenkin:2000jqa}:
\begin{align} \label{thicknesses}
r_{s}&\approx m_{s}^{-1}=(\sqrt{\lambda}\eta)^{-1}\\
r_{v}&\approx m_{v}^{-1}=(\sqrt{2}q\eta)^{-1}.
\end{align}
We choose our parameters to correspond to the Bogomolnyi-Prasad-Sommerfield (BPS) limit~\cite{Bogomolny:1975de,Prasad:1975kr}, where $r_s=r_v$. The energy per unit length of such a string is $\mu_{s}= 2\pi \eta^{2}$ \cite{Vilenkin:2000jqa}.

The symmetron model was originally introduced in the context of dark energy, with the symmetry breaking taking place at redshift $z \sim 1$, with the model parameters chosen so that the scalar field couples to matter with gravitational strength while satisfying the thin-shell condition in our galaxy. To estimate the tension of the resultant string, it helps, following \cite{Llinares:2014zxa}, to introduce the Compton wavelength of the scalar field in the vacuum, $\lambda_{0}=1/(\sqrt{2}\mu)$, as it relates $\mu$ to a cosmologically meaningful scale. We also introduce the dimensionless coupling $\beta =\eta M_{pl}/M^{2}$, and the matter density at symmetry breaking $\rho_{\rm SSB} = \mu^2M^2 = \rho_0/a_{\rm SSB}^3$, where $a_{\rm SSB}$ is the scale factor at symmetry breaking and $\rho_0$ is the matter density today, at $a=1$. Taking $\beta=1$, $a_{\rm SSB} = 0.5$, and $\lambda_0=0.1$ Mpc, gives a dimensionless string tension
\be
G\mu_{s}=2G\pi \eta^{2}=\frac{\lambda_{0}^{4}\beta^{2}\rho_{SSB}^{2}}{M^{2}_{p}} \approx 10^{-18}.
\ee
The string tension determines the metric deficit angle created by the string~\cite{Vilenkin:2000jqa}, and is a measure of the string's gravitational impact that can be compared to a typical gravitational potential $\Psi$ of an astrophysical object. At the surface of a star like our Sun, it is $\Psi/c^2 \sim 10^{-6}$, comparable to that at the edge of the Milky Way, while galaxy clusters and the cosmic web have $\Psi/c^2 \sim 10^{-5}$, comparable to the primordial metric fluctuations. These are much larger than the distortion of the metric created by a symmetron string. The cumulative effect of a moving string would, of course, be more pronounced, as it generates a wake behind it, but the expectation is that this too would not be observable for symmetron with a close-to-gravitational-strength coupling to matter. The ratio of the gravitational potential at the surface of a mass that has accreted onto a loop to the surface potential of a typical star like sun is $\Psi_{loop,string}/\Psi_{stellar} \sim  10^{-12}$, where we have considered a spherical mass that has accreted unto the loop. The same holds true for the wake of a straight string, as the surface potential of a cylindrical mass caused by a string is proportional to the string tension~\cite{Vilenkin:2000jqa}.

\section{Simulating symmetron cosmic strings interacting with matter halos}
\label{s:methods}

We numerically evolve Eqs.~(\ref{eq:EoM_1}) and (\ref{eq:EoMs_2}) in the temporal gauge, $A_0=0$, for several initial configurations of cosmic strings and matter density regions (halos) using the Crank-Nicholson method with two iterations~\cite{Teukolsky:1999rm}, in a $400^3$ simulation box with periodic boundary conditions (PBC). We adapt the numerical code developed in~\cite{Matsunami:2019fss}, originally used to simulate Abelian-Higgs-model (Nielsen-Olesen) strings, to the symmetron model. 

In what follows, we describe configurations involving straight strings and circular string loops interacting with spherical halos of uniform matter density. As a first step, we obtain field profiles for a straight string, a string loop, and a spherical matter halo separately. These are then combined into initial configurations in which the components are sufficiently far apart that their interactions are negligible at the initial time.

\subsection{Straight cosmic strings}

To start, we find the static straight string solution by substituting the Nielsen-Olesen anzatz given by Eqs.~(\ref{eq:str_scalar_ansatz}) and (\ref{eq:str_gauge_ansatz}) into the EOM and solving for functions $f(r)$ and $\alpha(r)$ using the relaxation method. The field configuration for a straight string moving in a given direction is obtained from the static solution via the Lorentz transformation.

A single straight string cannot satisfy PBC on faces of the simulation box that are parallel to the string axis, because the phase of the scalar field winds in opposite directions on opposing boundaries. To circumvent this, we include a second, parallel string of opposite winding, an anti-string, given by $\phi_{\rm as} = \bar{\phi}{\rm s} = \eta f e^{-i\theta}$. The combined configuration for a well-separated string-anti-string pair is then expressed as~\cite{Vilenkin:2000jqa}
\begin{align}
\Phi_{\rm pair}&=\frac{1}{\eta}\phi_{\rm s}\phi_{\rm as}  \label{eq:str_scalar_total}\\
\vec{A}_{\rm pair}&=\vec{A}_{\rm s}+\vec{A}_{\rm as} \label{eq:str_gauge_total}.
\end{align}

Even with this configuration, the total phase of the field generally fails to satisfy PBCs in a finite box, since each boundary point lies at a different distance and angle relative to the two strings. To address this, we impose a transition function that gradually suppresses the imaginary part of the field near the boundaries while preserving its magnitude, thereby reducing phase differences in those regions. Explicitly, we define
\ba
{\rm Im}(\tilde \Phi) &=& T(r) {\rm Im}(\Phi), \nonumber \\
{\rm Re}(\tilde \Phi) &=& {{\rm Re}(\Phi) \over |{\rm Re}(\Phi)|} \sqrt{{\rm Re}(\Phi)^{2}+{\rm Im}(\Phi)^2-{\rm Im}(\tilde \Phi)^2}
 \label{eq:modified}
\ea
where the transition function $T(r)$ is given by
\begin{equation}\label{eq:transition}
T(r)={1 \over 2}[1-\tanh((r-r_0)/w],
\end{equation}
with $r=\sqrt{x^2+y^2}$, $r_{0}$ defining the radius of the transition region, and $w$ its width. 

The same transition function is also applied to suppress the transverse components of the gauge field, $A_x$ and $A_y$, near the boundaries. We place the string configurations of interest within a cylindrical region defined by $r < r_0$, ensuring they remain well separated from the boundary and minimally affected by the transition. The field gradients introduced by this procedure generate a small amount of radiation within the simulation box, but this does not impact the qualitative features of the results we study.

\subsection{Loops}

\begin{figure}[tbp]
	\centering
	\includegraphics[width=.4\textwidth]{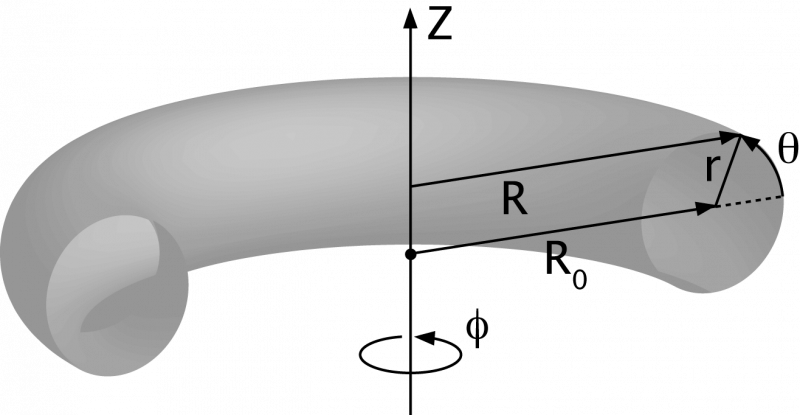}
	\caption{Poloidal-toroidal coordinate system. The coordinate $r$ is measured from the center of a circular cross-section of a torus with major radius $R_0$. The angle $\phi$ is the azimuthal angle around the loop, while $\theta$ defines the angle that $r$ makes with a reference axis in the $\phi = \text{constant}$ plane.}
	\label{fig:poloid}
\end{figure}

To simulate a loop, we implement the ansatzes in Eqs.~(\ref{eq:str_scalar_ansatz}) and~(\ref{eq:str_gauge_ansatz}) in a poloidal-toroidal coordinate system~\cite{Helfer:2018qgv}. The relation between Cartesian coordinates and poloidal-toroidal coordinates is given by
\begin{align}
x &= \cos(\phi)\left(R_{0} + r\cos\theta\right), \label{eq:trans_x} \\
y &= \sin(\phi)\left(R_{0} + r\cos\theta\right), \label{eq:trans_y} \\
z &= r\sin\theta, \label{eq:trans_z}
\end{align}
where $R_0$ is the radius of the circle defining the core of the loop ({\it i.e.}, the line of vanishing VEV), and $(r, \theta)$ are polar coordinates in the cross-sectional plane labeled by the toroidal angle $\phi$, as illustrated in Fig.~\ref{fig:poloid}.

Enforcing PBCs presents the same challenges as in the case of straight strings. Therefore, we apply a similar boundary-smoothing procedure with two key modifications: (i) the transition function in Eq.~(\ref{eq:transition}) is now defined in terms of the spherical radial coordinate, $r = \sqrt{x^2 + y^2 + z^2}$; (ii) the scalar field is multiplied by the transition function (\ref{eq:transition}), which gradually brings it to zero outside a spherical region, while simultaneously adding a compensating term that transitions from zero to the uniform VEV at the same rate. This ensures that the field approaches the vacuum expectation value smoothly at large distances from the loop.

\subsection{Adding mass halos}

In this work, we consider spherical matter overdensities with a density profile given by
\begin{equation}
\rho(x, y, z) = \kappa \left[1 - \tanh\left(\nu (r - \chi)\right)\right], \label{eq:mass}
\end{equation}
where the parameters $\kappa$, $\nu$, and $\chi$ control the amplitude, steepness, and location of the transition, respectively. Here, $r = \sqrt{(x - x_m)^2 + (y - y_m)^2 + (z - z_m)^2}$ is the radial distance from the center of the mass halo, located at coordinates $(x_m, y_m, z_m)$.

To determine the scalar field profile around an isolated mass halo, we adopt a spherically symmetric static ansatz with the gauge field set to zero and a uniform scalar field phase (which we set to zero without loss of generality). We substitute $|\phi| = \phi_m(r)$ and $\vec{A}_m = 0$ into the equations of motion and solve for $\phi_m(r)$ using the numerical relaxation method. This initial configuration is an exact solution, as the equations of motion have no source term that can generate gauge fields or spatial dependence of the phase. 

To construct the combined initial configuration consisting of a straight string (or a loop) and a mass halo, we take them to be separated by a distance large enough for the scalar field amplitude gradients and the potential energy density to vanish in between\footnote{Because of the non-linearity of the scalar field potential, a superposition of two solutions of the equations of motion is not, in general, a valid solution. However, the superposition works in the limit where the localized solutions are well-separated and the potential energy density vanshes in between. In practice, in numerical applications, the distance does not need to be larger than a couple of string core sizes, as any residual field gradients radiate away.}. We then multiply the scalar fields of the individual solutions and add the gauge fields. That is,
\begin{equation}
\Phi(x, y, z) = \phi_m(r)\, \frac{\Phi_{\text{string}}(x, y, z)}{\eta}, \ \vec{A}=\vec{A}_{\text{string}},\label{eq:combined}
\end{equation}
where $\Phi_{\text{string}}$ and $\vec{A}_{\text{string}}$ are the complex scalar field and the gauge field of the isolated string solution, and $\eta$ is the vacuum expectation value. In the case of an isolated static straight string, at a distance far from the string, the scalar amplitude is constant and equal to the VEV, while the phase gradients and the gauge field cancel each other to set the covariant derivative to zero. This cancellation remains unaltered in the combined ansatz for a static string in the large separation limit. For a moving string or the loop at a finite separation from the halo the field gradients due to the departure of the initial configuration from an exact solution radiate away before the string and the halo get close. We adopt this approach to assemble composite initial configurations involving multiple halos.

\section{Results and Analysis}
\label{s:results}

\begin{figure*}[htbp!] 
	\includegraphics[width=0.95\textwidth]{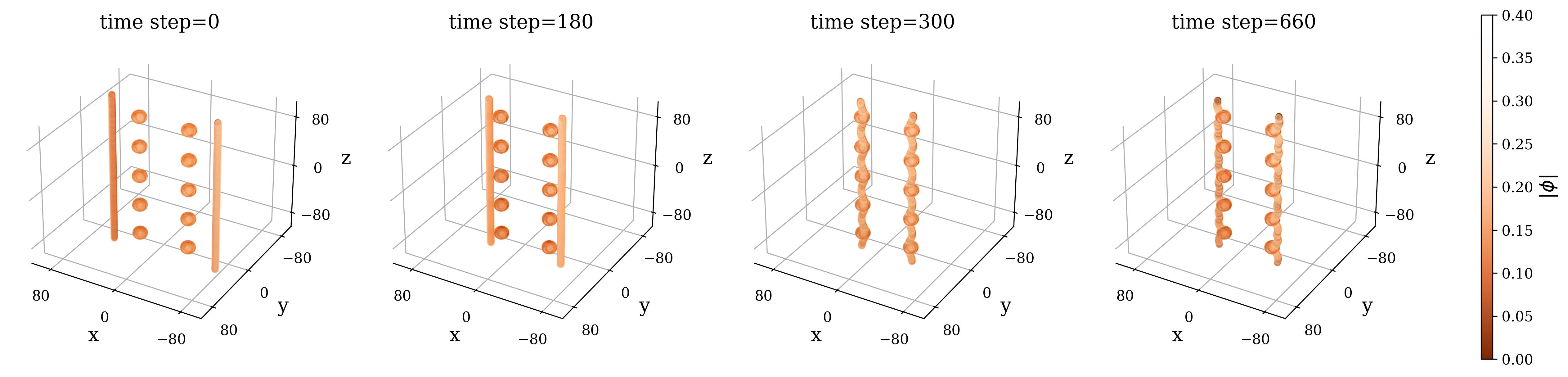}
	\includegraphics[width=0.95\textwidth]{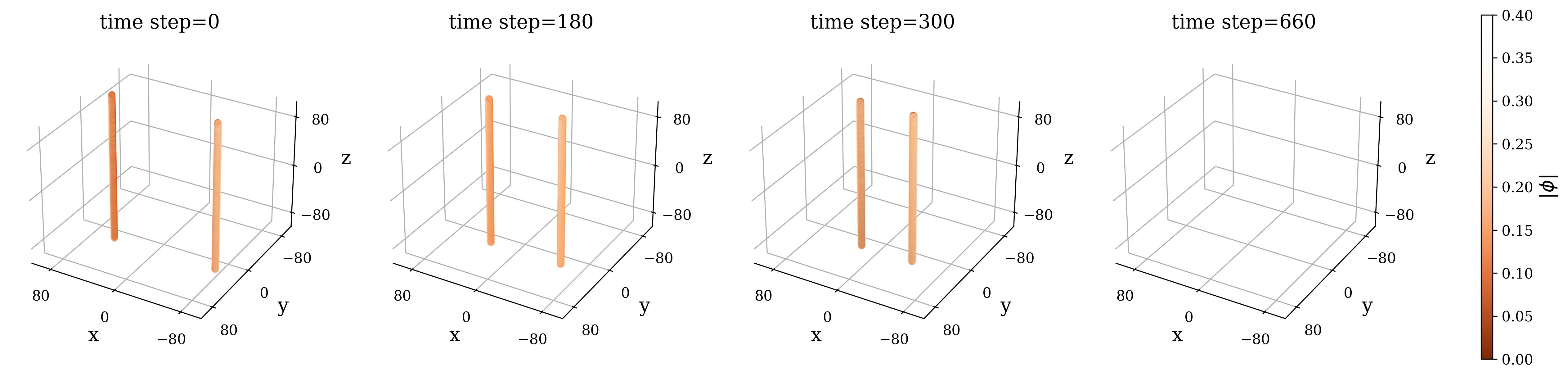}
	\caption{Top: A pair of a string and an anti-string shown at four time snapshots: at the beginning, as they approach the mass halos, and at two later times when both are pinned to the halos. Bottom: The same pair shown at corresponding time steps, but without the halos. The strings reach each other and annihilate. In both cases, the initial velocity of the strings is $v = 0.5$. The simulation parameters are: $\eta = 1.0$, $q = 1.0$, $\lambda = 0.5$, $M = 0.5$, $\kappa = 3.0$, $\nu = 30.0$, $\chi = 2.5$.}
	\label{fig:strings}
\end{figure*}

\begin{figure*}[htbp!] 
	\centering	
	\includegraphics[width=0.95\textwidth]{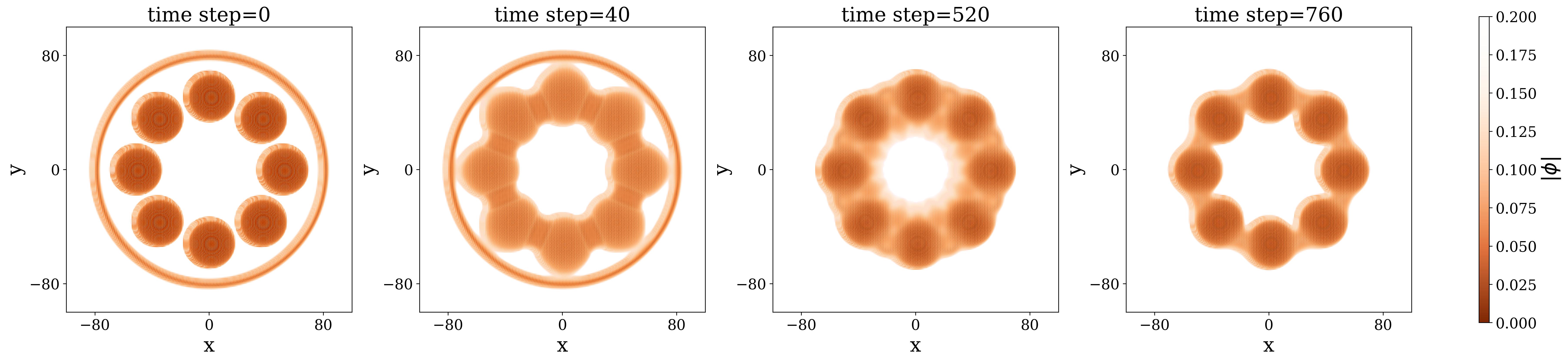}
	\includegraphics[width=0.95\textwidth]{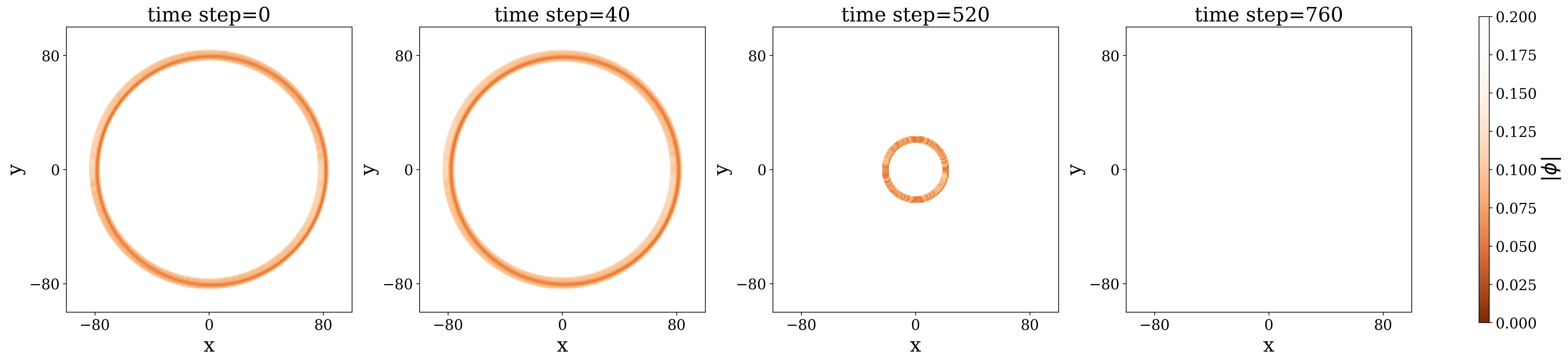}
	\caption{Top: A circular loop with eight mass halos arranged in a ring is shown at four time snapshots: initially, just before reaching the halos, and at two later times when the loop becomes pinned to the masses. Bottom: The same loop configuration without halos, shown at corresponding times. The loop contracts and annihilates. Simulation parameters: $\eta = 0.5$, $q = 1.0$, $\lambda = 0.5$, $M = 0.5$, $\kappa = 3.0$, $\nu = 10.0$, $\chi = 8.0$.}
	\label{fig:loop}
\end{figure*}

In our simulations, we set the symmetron model parameters to $\eta = 1.0$, $q = 1.0$, $\lambda = 0.5$ and $M = 0.5$, corresponding to the BPS-limit string solutions. The mass halo parameters in Eq.~(\ref{eq:mass}) were set to $\kappa = 3.0$, $\nu = 30.0$ and $\chi = 2.5$ for the straight string demonstrations, and $\kappa = 3.0$, $\nu = 10.0$, $\chi = 8.0$ for the loop simulations. We choose the lattice spacing to be $\Delta x = 0.5$, which allows for approximately $10$ lattice points inside the string core, and $8$ ($24$) lattice points inside the ``thin-shell'' of the scalar field profile around the mass halos in the straight string (loop) simulations. The extensive tests of our code performed in \cite{Matsunami:2019fss} have shown that the accuracy of the total energy conservation was a good indicator of numerical convergence. We tested that the energy was conserved to better than $5$\% for all configurations evolved, well-sufficient for the qualitative demonstrations of the expected string-halo interactions.

Our simulations confirm the expectation that strings preferentially attach themselves to mass halos. We demonstrate that this behavior can stabilize otherwise unstable string configurations, including both cosmic string loops and straight string-anti-string pairs. The results are shown in Figs.~\ref{fig:strings} and \ref{fig:loop}, where we plot the absolute value of the scalar field. The model and simulation parameters are provided in the figure captions.

Fig.~\ref{fig:strings} shows a string and an anti-string moving toward each other with initial velocity\footnote{We give the strings an initial velocity to accelerate the annihilation process. String and anti-string pairs naturally attract each other, but for local gauge strings the interaction range is exponentially suppressed, making the timescale for annihilation too long for direct numerical observation.} $v = 0.5$. In the absence of mass halos, as shown in the lower panel, the strings meet and annihilate. In contrast, the upper panel shows that when ten mass halos are placed along the strings' trajectory, the strings become pinned to the halos and remain stationary, forming a stable configuration.

Fig.~\ref{fig:loop} illustrates the evolution of a circular string loop under similar conditions, with and without eight mass halos arranged along a smaller, concentric circle. Loops are inherently unstable due to the absence of topological charge and, as the lower panel shows, quickly shrink and decay into radiation. However, in the presence of halos (upper panel), the loop attaches to the masses and remains stabilized.

Let us now elaborate on why it is energetically favorable for strings to attach themselves to matter halos, and under what conditions loops or straight strings can be stabilized. This is most easily understood in the Nambu-Goto limit, where the string is much thinner than the width of the halo and its energy is dominated by tension $\mu_s$, which is the same as the energy per unit length. In this regime, a circular loop of radius $R$ has total energy $E = 2\pi \mu_s R$. Now consider the case where segments of the loop pass through $N$ mass halos. When the thin-shell condition (\ref{eq:thin-shell}) is satisfied, the scalar field inside the halo vanishes, thus matching the value in the string core and rendering the string effectively massless in those regions. As a result, the segments lying inside the halos contribute negligibly to the total energy, and the loop energy is reduced to $E = 2\pi \mu_s R - N \mu_s \ell$, where $\ell$ denotes the length of the loop contained within each halo. This approximation is valid provided that (i) the thin-shell condition holds, and (ii) the string thickness is much smaller than the halo radius, ensuring that the string is fully embedded within the screened region.

\begin{figure}[tbp!]
	\centering
	\begin{minipage}{0.22\textwidth}
		\centering
		\includegraphics[width=\textwidth]{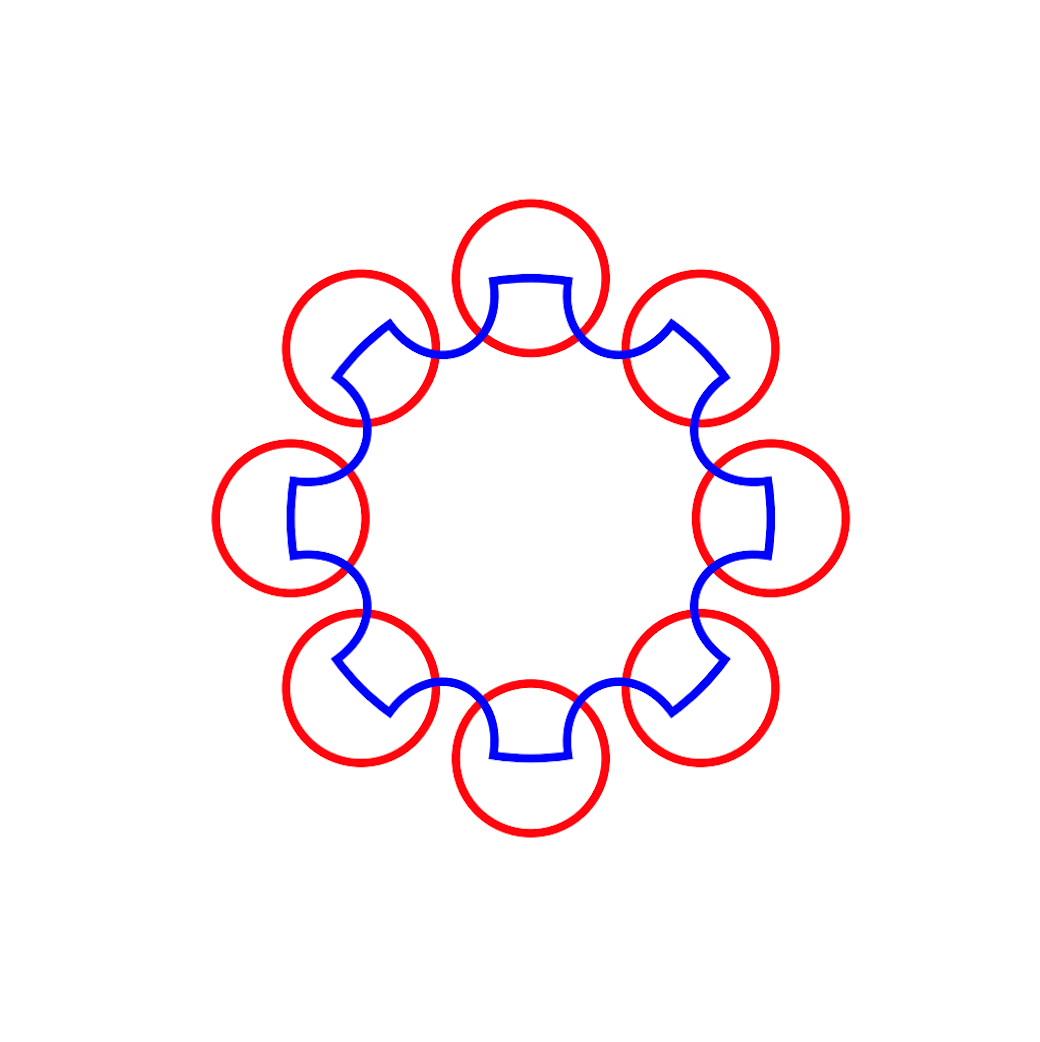}
	\end{minipage}
	\hspace{0cm}
	\begin{minipage}{0.22\textwidth}
		\centering
		\includegraphics[width=\textwidth]{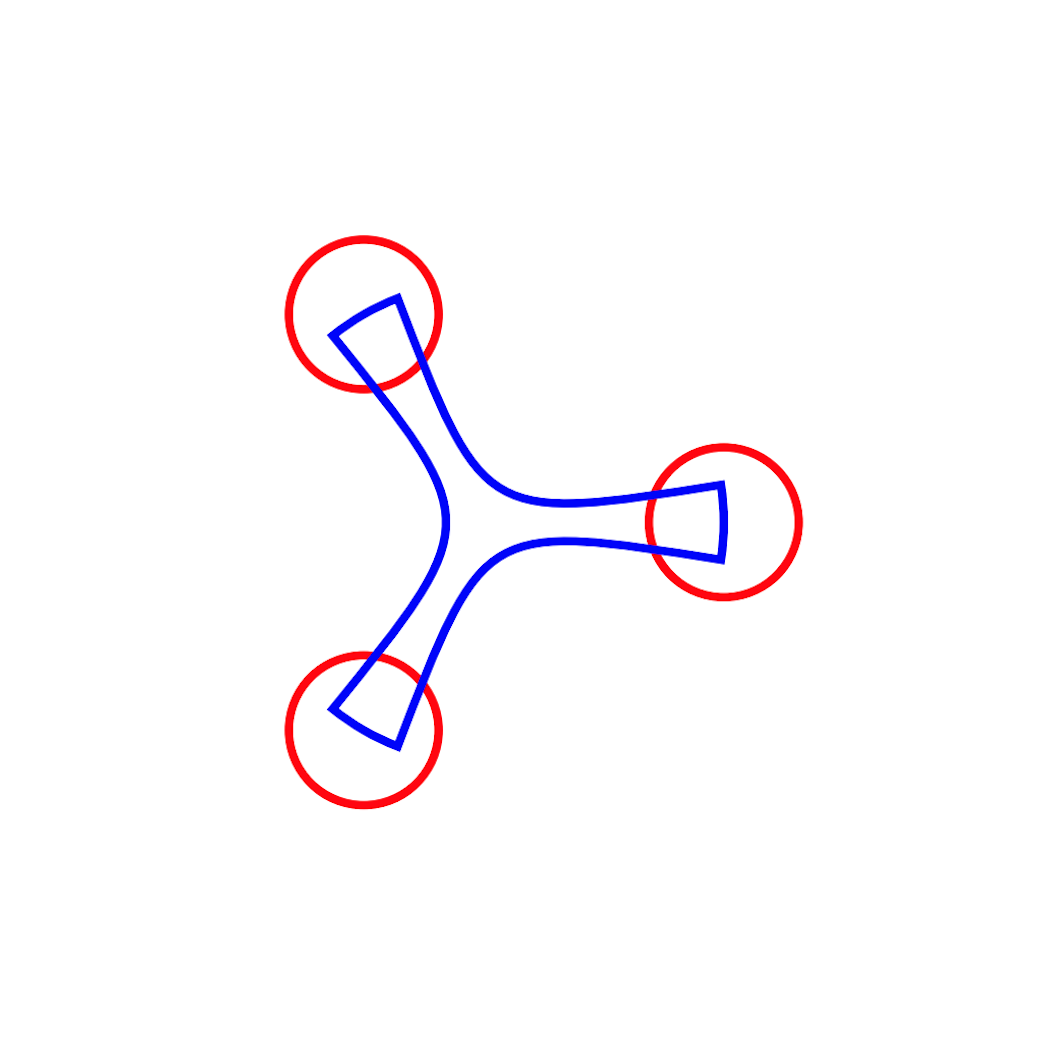}
	\end{minipage}
	\caption{A schematic illustration of how unpinned segments of a loop behave after encountering mass halos. Left: The number and spacing of halos prevent the unpinned segments from reconnecting; they oscillate around straight-line configurations. Right: If the segments carry sufficient kinetic energy, they bend and reconnect, leading to loop decay.}
	\label{fig:bending}
\end{figure}

More generally, the loop may deform and lose its circular symmetry. In that case, the total energy becomes
\be
E = \mu_s L - N \mu_s \ell,
\label{eq:loop_general_mass}
\ee
where $L$ is the invariant length of the loop~\cite{Vilenkin:2000jqa}. Energy minimization then corresponds to minimizing the total string length outside the halos. In the absence of halos, the minimum-energy configuration corresponds to $L \to 0$, {\it i.e.}, a collapsing loop. When matter halos are present, segments of the loop become pinned, while the unpinned segments adjust to minimize energy subject to the constraints of initial momentum and boundary conditions. Even a loop initialized at rest gains momentum as it contracts. Upon encountering the halos, it retains this kinetic energy, allowing the unpinned segments to bend. These segments may oscillate, remain pinned in a quasi-stable configuration, or, if sufficiently energetic and favorably spaced, reconnect and annihilate. Fig.~\ref{fig:bending} illustrates these possible outcomes.

A similar argument applies to straight strings. For a string of length $L$, with $N$ segments embedded in halos, the energy is given by Eq.~(\ref{eq:loop_general_mass}). Thus, straight strings may also become pinned if the unembedded segments lack the momentum needed to reach and annihilate with an anti-string. The stabilization of such configurations depends on the number and arrangement of the halos, as well as the initial conditions and motion of the string.

\

\section{Summary and Discussion}
\label{s:summary}

In this work, we have demonstrated that cosmic strings in the complex symmetron model preferentially attach to matter halos and can form stabilized configurations, including string loops and string-anti-string pairs, that would otherwise collapse or annihilate. This behavior arises naturally from the screening mechanism in the symmetron model, which suppresses the string's energy inside high-density regions.

Previous studies investigated scattering and capture of minimally coupled cosmic strings by black holes~\cite{DeVilliers:1998nm,Snajdr:2002aa,Dubath:2006vs,Deng:2023cwh,Bambhaniya:2024hzb} and their interactions with matter in galaxies and clusters~\cite{Gasilov:1985}. While we have not considered the backreaction of non-minimally coupled strings or domain walls on the distribution of matter, this remains an intriguing direction for future investigation, particularly in the context of dark matter and dark sector interaction models. In scalar-tensor theories with universal coupling, such as the symmetron, the requirement of screening in high-density environments constrains the allowed tension of topological defects, typically limiting their observable gravitational effects. However, if the defects exist solely in the dark sector, these constraints are significantly relaxed. One could, in principle, have heavier strings and domain walls composed entirely of dark-sector fields, which might still leave an imprint on the evolution and morphology of dark matter structures.

We also note that the phenomenology of symmetron defects bears a striking resemblance to that of vortices in superfluids and domain walls in ferromagnets, which are known to interact strongly with their environment. In particular, such defects tend to pin themselves on inhomogeneities or impurities in the medium~\cite{PhysRevB.98.054406, PhysRevX.14.041039}, much like symmetron strings and domain walls pin themselves to matter halos. This analogy may provide a useful framework for understanding the dynamics of screening-stabilized topological defects in cosmological settings.

\

{\bf Acknowledgments.} We thank Tanmay Vachaspati for helpful discussions. This research was enabled in part by support provided by the BC DRI Group and the Digital Research Alliance of Canada ({\tt alliancecan.ca}). A.N and L.P. are supported in part by the National Sciences and Engineering Research Council (NSERC) of Canada.


%

\end{document}